\documentclass[aps,pra,twocolumn,floatfix,nofootinbib]{revtex4}
\usepackage{amsfonts}
\usepackage{amsmath}
\usepackage{amssymb}
\usepackage{epsfig}
\usepackage{bm}

\usepackage{graphicx}

\begin{document}

\title{Beyond sub-Rayleigh imaging via high order correlation of speckle illumination}

\author{Fu Li$^1$, Charles Altuzarra $^1$, Tian Li$^{1,2}$ and M.O. Scully $^{1,3,4}$, G.S. Agarwal $^{1,2}$ }%

\address{$^1$ Institute for Quantum Science and Engineering, Texas A$\&$M University, College Station, TX 77843, USA}
\address{$^2$ Department of Biological and Agricultural Engineering, Department of Physics and Astronomy, Texas A$\&$M University, College Station, TX 77843, USA}
\address{$^3$ Department of Mechanical and Aerospace Engineering, Princeton University, Princeton, NJ 08544, USA}
\address{$^4$ Quantum Optics Laboratory, Baylor Research and Innovation Collaborative, Waco, TX 76704, USA}

\begin{abstract}
Second order intensity correlations of speckle illumination are extensively used in imaging applications that require going beyond the Rayleigh limit. The theoretical analysis shows that significantly improved imaging can be extracted from the study of increasingly higher order intensity cumulants. We provide experimental evidence by demonstrating resolution beyond what is achievable by second order correlations. We present results up to $20^{th}$ order. We also show an increased visibility of cumulant correlations compared to moment correlations. Our findings clearly suggest the benefits of using higher order intensity cumulants in other disciplines like astronomy and biology. \noindent{\it sub-Rayleigh imaging, high order correlation, speckle illumination, cumulant, moment\/}
\end{abstract}

\maketitle


\section{Introduction}

Speckles in optics are well known and have various applications. Typically, speckles are produced in the scattering of a coherent beam of light by a random medium \cite{goodman2007speckle,goodman1976some,dunn2001dynamic,Dainty1975laser,gatti2008progress,Devaux_2016}. The propagation difference of partially coherent beams and coherent lasers through a random medium has been successfully explained by Dogariu et. al \cite{dogariu2003propagation,shirai2003mode,siviloglou2007observation}. The medium scatters a coherent beam in various directions with randomly varying phases. The scattering medium under fairly general conditions produces fields that can be modeled as Gaussian fields. The characteristics of the medium can be extracted from the spatial and temporal coherence of the medium. The spatial coherence information is in turn obtained from the intensity-intensity correlations \cite{baleine2006correlated,gatti2004ghost,valencia2005two}. Such studies provide a wealth of information on the medium \cite{lahiri2009determination}. Speckled illumination is produced using a rotating ground glass \cite{arecchi1965measurement} and has been shown to beat the diffraction limit \cite{classen2017superresolution,oh2013sub,wang2015spatial,smith2018turbulence,sprigg2016super}. Second order intensity correlations beat the diffraction limit by a factor of $\sqrt{2}$. Most studies on beating the diffraction limit use second order intensity correlations. Also the super-resolution optical fluctuation imaging (SOFI) technique based on cumulant correlations, is limited to samples with particular intrinsic blinking characteristics~\cite{dertinger2009fast}. Differently, in this work, super-resolution does not rely on the object's blinking characteristics, but instead intensity fluctuations from the speckled light source are used with high order correlations, which results in super-resolution imaging beyond the Rayleigh limit. Several other theoretical works have investigated the uses of higher order correlations~\cite{vigoren2018optical,oppel2014directional,chan2009high}.\\

In this letter we experimentally demonstrate higher order correlations for imaging applications by using speckle illumination, and analyze the reliability of high order correlations, which consolidates the potential for applications. We first present a theoretical analysis based on the cumulants of the measured intensity distributions and show that such cumulants can beat the diffraction limit by a factor of $\sqrt{n}$. We then present experimental confirmation. Our results with speckle light underline two advancements: 1) the improvement in resolution generated by moment correlation orders $N > 2$, and 2) the superior resolution offered by cumulants as compared to moments. In addition, our experiment determines that speckle illumination is also valid for non-fluorescent samples, which is very important for label free bio-imaging.
\begin{figure*}[htbp]
\centering
\includegraphics[width=\linewidth]{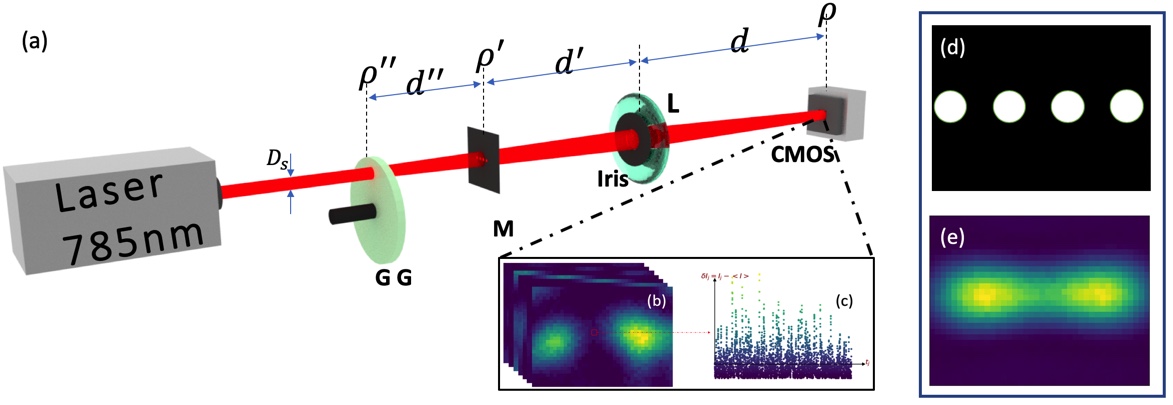}
\caption{Color online. The experimental setup.  (a) The laser incident on a rotating ground glass (GG) generates speckled light. The light transmitting through the mask (M) and iris is collected by the camera (CMOS) with a lens (L) f=150mm. Here, distance $d^{'}=d=300mm$. (b) Several image frames are recorded to compute the high order correlation images. (c) A pixel’s temporal intensity fluctuation.  (d) The mask object, notably four dots. (e) Laser illumination intensity (Int.) imaging without Ground Glass (GG), for which the image of the four dots is blurred since the Rayleigh limit is twice the distance of the dot separation.}
\label{fig:1}
\end{figure*}

\section{Theoretical Basis}
\label{sec:Theoretical}

As illustrated in figure.1, the experimental setup uses a continuous wave laser of 785nm in wavelength (Omicron LuxX-785) incident on a rotating ground glass to generate a speckle pattern. This source is used to illuminate a micro-fabricated mask object (Toppan Digidat), notably four holes milled in a chromium layer deposited on glass, for which their diameter $l_0$ equals their separation of $l= 25\mu m$. An aperture limited by an adjustable iris and an $f=150mm$ imaging lens (L), produce an image with a magnification factor $M = d^{'}/d = 1$ with $d^{'}= d = 300mm$, where $d^{'}$ and d are the mask object distance and image distance from the lens, respectively. Here the image is mapped on a 1280x1024 pixels’ CMOS camera (Thorlabs DCC3240N), whose pixel area is 5.3 um x 5.3 um.

In our setup, the electric field generated from the ground glass propagates to a transmission mask with the transmission coefficient $t\left(\overrightarrow{\rho}_{i}^{'}\right)$, where $\overrightarrow{\rho}_{i}^{'}$ is a position on the mask object plane. For the transmitted electric field, the intensity fluctuation correlation, which is crucial to surpass the Rayleigh limit, is described by \cite{sprigg2016super,shih2016introduction}
\begin{equation}
\begin{aligned}
\left\langle \delta I\left(\overrightarrow{\rho}_{1}^{'}\right)\delta I\left(\overrightarrow{\rho}_{2}^{'}\right)\right\rangle
&=\left|G^{(1)}\left(\overrightarrow{\rho}_{1}^{'},\overrightarrow{\rho}_{2}^{'}\right)\right|^{2} \\
&\propto \left|t\left(\overrightarrow{\rho}_{1}^{'}\right)\right|^{2}\left|t\left(\overrightarrow{\rho}_{2}^{'}\right)\right|^{2}e^{-\frac{\left(\overrightarrow{\rho}_{1}^{'}-\overrightarrow{\rho}_{2}^{'}\right)^{2}}{l_{c}^{2}}},
\end{aligned}
\label{eq:1}
\end{equation} with intensity fluctuation $\delta I\left(\overrightarrow{\rho}_{i}^{'}\right)=I\left(\overrightarrow{\rho}_{i}^{'}\right)-\left\langle I\left(\overrightarrow{\rho}_{i}^{'}\right)\right\rangle$ and a coherence length $l_c \propto \lambda d^{''}/D_s$. 
Here, $G^{(1)}\left(\overrightarrow{\rho}_{1}^{'},\overrightarrow{\rho}_{2}^{'}\right)=\langle E^*(\overrightarrow{\rho}_{1}^{'}) E(\overrightarrow{\rho}_{2}^{'})\rangle$ is the first order correlation~\cite{goodman2007speckle}. The electric field generated from the ground glass propagates to the mask, the transmission part of electric field at the $\overrightarrow{\rho}^{'}$ plane is given by $E\left(\overrightarrow{\rho}^{'}\right)=t\left(\overrightarrow{\rho}^{'}\right)\int G\left(\overrightarrow{\rho}^{'}-\overrightarrow{\rho}^{''},d^{''}\right)E\left(\overrightarrow{\rho}^{''}\right)d\overrightarrow{\rho}^{''}$
, where $G\left(\overrightarrow{\rho}^{'}-\overrightarrow{\rho}^{''},d^{''}\right)\backsimeq\frac{e^{ikd^{''}}}{i\lambda d^{''}}e^{\frac{i\pi}{\lambda d^{''}}\left(\overrightarrow{\rho}^{'}-\overrightarrow{\rho}^{''}\right)^{2}}$~\cite{oh2013sub,wang2015spatial} is the Fresnel propagator. With the assumption of $\left\langle E\left(\overrightarrow{\rho}_{1}^{''}\right)E\left(\overrightarrow{\rho}_{2}^{''}\right)\right\rangle \sim \delta\left(\overrightarrow{\rho}_{1}^{''}-\overrightarrow{\rho}_{2}^{''}\right)$ at the ground glass plane, $ G^{(1)}\left(\overrightarrow{\rho}_{1}^{'}, \overrightarrow{\rho}_{2}^{'}\right) \propto t^{*}\left(\overrightarrow{\rho}_{1}^{'}\right)t\left(\overrightarrow{\rho}_{2}^{'}\right)e^{-\frac{\left(\overrightarrow{\rho}_{1}^{'}-\overrightarrow{\rho}_{2}^{'}\right)^{2}}{l^{2}}}$ is obtained, see \ref{Correlation calculation}.

Thus, when the mask’s hole separation is approximately equal to or larger than the coherent length ($l_c$), the holes contribute independently and randomly, $\left\langle \delta I\left(\overrightarrow{\rho}_{i}^{'}\right)\delta I\left(\overrightarrow{\rho}_{j}^{'}\right)\right\rangle _{i\neq j}\simeq0$
, see Eqn.(\ref{eq:1}). The holes are imaged onto the camera, whose intensity $I\left(\overrightarrow{\rho}\right)$ at $\overrightarrow{\rho}$ position is formulated by 

\begin{equation}
I\left(\overrightarrow{\rho}\right)=\sum_{i=1,2,3,4}h_{i}I_{i},
\label{eq:2}
\end{equation} where $h_i=h\left(\overrightarrow{\rho} - \overrightarrow{\rho}_i \right)$ is the point-spread function (PSF) of the imaging system and $I_i=I\left(\overrightarrow{\rho}_i^{'} \right)$, and the summation is over all points in the holes. The width of the PSF gives the Rayleigh limit of an imaging system, which blurs out any point-like object to an Airy disk. Instead of using intensity imaging, we use cumulants to demonstrate super-resolution. 

The quantity $I\left(\overrightarrow{\rho}\right)$ is a statistically fluctuating quantity. We evaluate cumulant generating function $K\left(\beta\right)$ of $I\left(\overrightarrow{\rho}\right)$

\begin{equation}
\begin{aligned}
K\left(\beta\right) & =ln\left\langle exp\left(\beta I\left(\overrightarrow{\rho}\right)\right)\right\rangle \\
&=\sum_{i}ln\left\langle exp\left(\beta h_{i}I_{i}\right)\right\rangle,
\end{aligned}
\label{eq:3}
\end{equation} where $I_i$ is the statistical independence of the variables. We can rewrite Eqn.(\ref{eq:3}) in terms of the cumulants $\kappa_n$ of $I\left(\overrightarrow{\rho}\right)$ and $\kappa_{in}$ of  $I\left(\overrightarrow{\rho}_i\right)$,

\begin{equation}
\begin{aligned}
\kappa_{n}\left(\overrightarrow{\rho}\right)=\sum_{i}\left(h\left(\overrightarrow{\rho}-\overrightarrow{\rho}_{i}\right)\right)^{n}\kappa_{in} .
\end{aligned}
\label{eq:4}
\end{equation}

Note that there are no cross terms in Eqn.(\ref{eq:4}).
In contrast, the $n^{th}$ order moment is given by

\begin{equation}
\begin{aligned}
m_{n}=\left\langle \left(I\right)^{n}\right\rangle =\left\langle \left(\sum_{i}h_{i}I_{i}\right)^{n}\right\rangle .
\end{aligned}
\label{eq:5}
\end{equation}

$m_n$ is composed of contributions from terms which are products of the PSF like $h_ih_j$ with ($i\neq j $). For example, the $4^{th}$ order moment is 

\begin{equation}
\begin{aligned}
m_{4}=\sum_{i}h_{i}^{4}\left\langle \left(I_{i}\right)^{4}\right\rangle +\sum_{i\neq j}h_{i}^{2}h_{j}^{2}\left\langle (I_{i})^{2}(I_{j})^{2}\right\rangle .
\end{aligned}
\label{eq:6}
\end{equation}

`Noisy' terms like the last term in Eqn.(\ref{eq:6}) do not appear in $\kappa_4$, therefore, imaging based on higher order cumulants $\kappa_n$ are much more effective than images based on $m_n$. 

In comparison with the intensity imaging given by $I\left(\overrightarrow{\rho}\right)=\sum_{i=1,2,3,4}h_{i}I_{i}$, Eq.(\ref{eq:4}) suggests that the $n^{th}$ order cumulant can yield imaging resolution improvement by a factor $\sim\sqrt{n}$, due to its narrowed effective PSF $h_{eff}=h^n$. By extension, with respect to a $G^{(2)}$ image~\cite{oh2013sub,wang2015spatial}, an improvement by a factor of $\sim\sqrt{n/2}$ can be achieved. To simplify the calculation, the cumulant formulas are expressed by the central moment form , see \ref{The cumulant formulas}.
\begin{align}
\kappa_{n}=\mu_{n}-\sum_{i=1}^{n-1}\left(\begin{array}{c}
n-1\\
i
\end{array}\right)\kappa_{n-i}\mu_{i}, \label{eq:7}
\end{align}
where $\mu_n =\left\langle \left(I - \left\langle I \right\rangle \right)^n\right\rangle $ is the central moment.

We next present experimental results based on $\kappa_n$ and $m_n$.

\section{Experimental Results}


In our work, as opposed to previous works in speckle imaging~\cite{oh2013sub,wang2015spatial}, we render the image completely indistinguishable by minimizing the aperture of the pinhole. The reason for excessively blurring the image is to demonstrate the power of this method that uses correlation orders of a factor 10 higher than what has been demonstrated so far experimentally with speckle light. Indeed though, our technique has an important requirement, each hole must be independently fluctuating in intensity. In our experiment, this condition is satisfied by engineering the speckles of the light source in relation to the object's structures. Thus, the coherence length of the source (i.e., the distance between speckles) needs to match the distance between the micro-structured holes on the object.

\begin{figure}[htbp]
\centering
\fbox{\includegraphics[width=\linewidth]{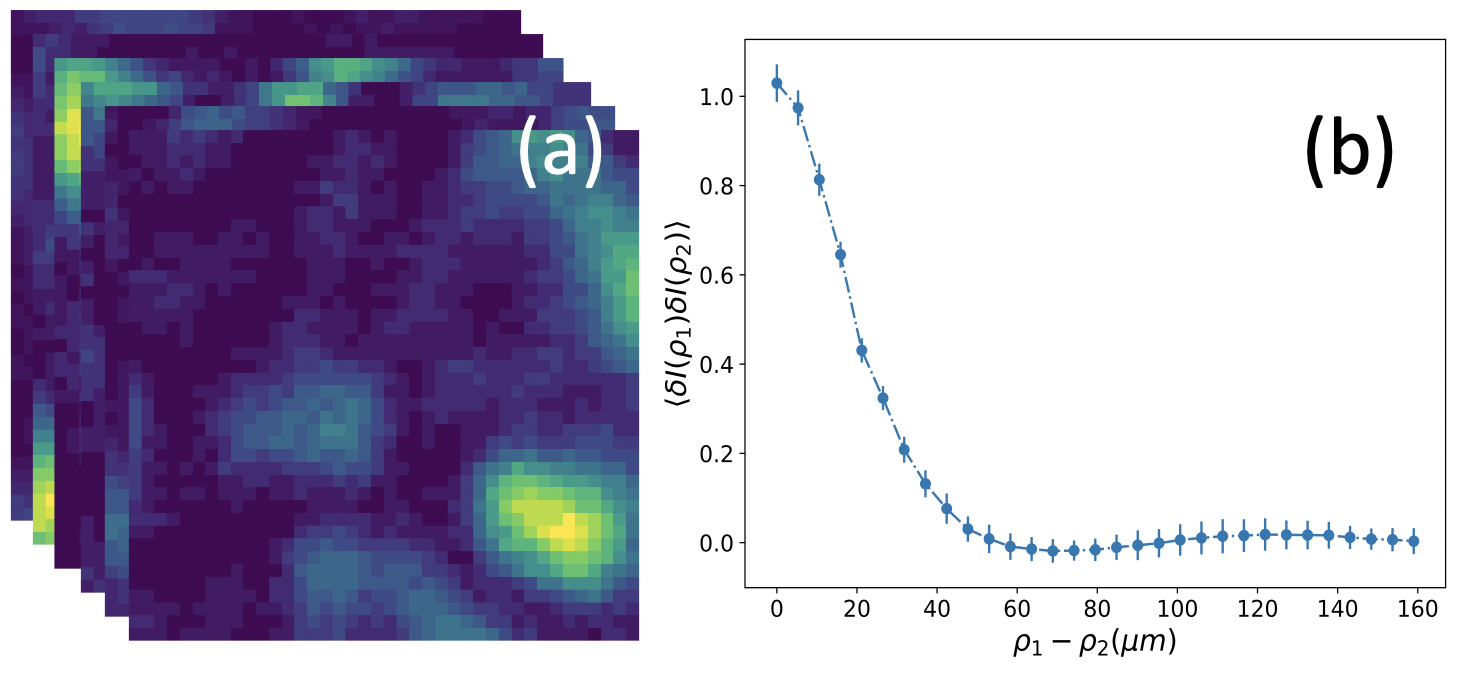}}
\caption{Color online. The speckle pattern coherence length $l_c \propto 24\mu m$. The target object is removed, and multi-frames images are measured, see (a). The correlations of the intensity fluctuation $\left\langle \delta I\left(\rho_{1}^{'}\right)\delta I\left(\rho_{2}^{'}\right)\right\rangle$ are calculated, see (b)}
\label{fig:7}
\end{figure}

In this section, our discussion starts with the generation of the speckle light source, which is a key point for understanding this experiment. Independent intensity fluctuations manifest themselves and are characterized by   $g^{(2)}\left(\overrightarrow{\rho}_{1}^{'},\overrightarrow{\rho}_{2}^{'}\right) \propto \left\langle \delta I\left(\overrightarrow{\rho}_{1}^{'}\right)\delta I\left(\overrightarrow{\rho}_{2}^{'}\right)\right\rangle$, where $\delta I$ is the intensity fluctuation at the target object plan. Thus, the measurement of the camera without the target object was produced, as shown in figure.\ref{fig:7} (a). These images were then used to calculate the coherence length, plotted in figure.\ref{fig:7} (b). In fact, based on the relation $l_c \propto \lambda d^{''}/D_s$, speckle pattern of different coherence lengths can be obtained, namely by changing the beam size $D_S$.


\begin{figure}[htbp]
\centering
\fbox{\includegraphics[width=\linewidth]{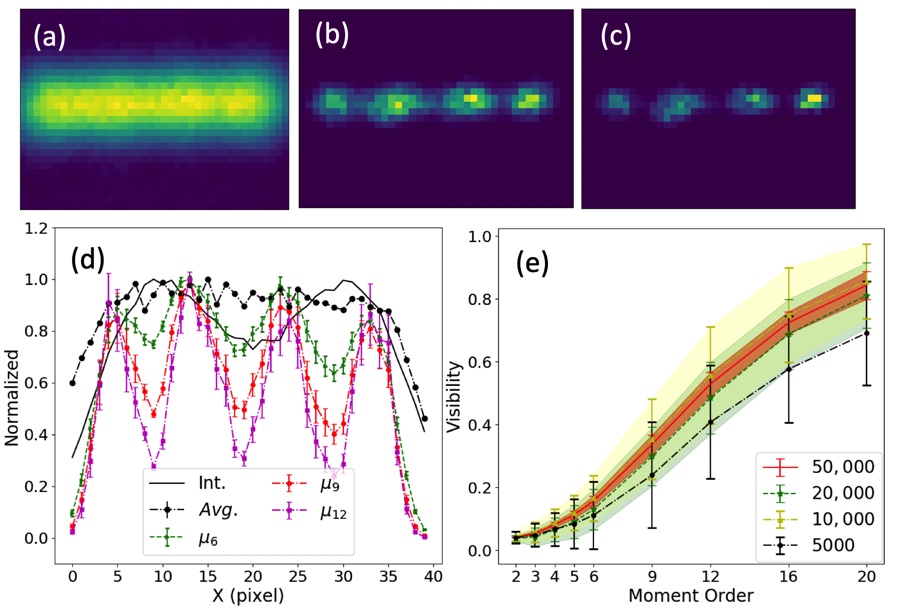}}
\caption{Color online. Comparison of traditional intensity imaging and high order moment imaging. (a) The average intensity (Avg.) imaging of speckle illumination, and the images reconstructed by the $12^{th}$, $20^{th}$ order central moment (b, c). (d) The contrast comparison from images of different orders by summing two rows of pixels that are centered with the holes. (f) The visibility and standard deviation as a function of the moment orders computed by using the different frame numbers.}
\label{fig:2}
\end{figure}

In our experiment, the coherence length is 24$\mu m$, which approximately equals to the separation 25$\mu m$. By obtaining the intensity fluctuation at each pixel and post-processing them, as illustrated in figure.\ref{fig:1} (b),(c), moment-generated images were produced in demonstrating that this technique provides resolution beyond the Rayleigh limit. This is shown, first and foremost, by completely blurring the image, which is done here by adjusting the aperture of the iris to a diameter of $5.75mm$. This size provides a Rayleigh limit of $\delta x  =0.61\lambda M/NA= 50\mu m$, which is two times the mask’s hole separation of $l = 25\mu m$. Thusly, it comes as no surprise that the laser illumination intensity image without Ground Glass (GG), portrayed in figure.\ref{fig:1} (e), yields only two peaks, and the average intensity of the speckle illumination image yields a completely blurred image as shown in figure.\ref{fig:2} (a). In contrast, the $12^{th}$ order central moment $\mu_{12}$ and the $20^{th}$ order central moment $\mu_{20}$ in figure. \ref{fig:2}  (b) and (c) respectively, result in a well-defined mask object. In fact, the Rayleigh limit is so large that only higher orders provide a definable image. Indeed, the difference in contrast is ascertained in figure.\ref{fig:2} (d). In this figure, the normalized photon counts are plotted in the horizontal direction, for two rows of pixels. As expected, a higher order moment provides a higher visibility image, and the four dots’ feature is recovered where intensity imaging yields a completely blurred image with no contrast. 

An imaging technique is unreliable if the results are not reproducible with high confidence. Indeed, the practicality of an imaging technique is qualified by the performance and reliability, which in mathematical language translates to high visibility $(I_{max}-I_{min})/(I_{max}+I_{min})$ and low standard deviation, respectively. Without loss of generality, we fix $I_{max}$ to be the second peak and the $I_{min}$ to be the second valley. For our setup, due to the randomness of the speckle illumination, it is clear that a large number of frames are required to compute a well-defined mask image. 

From figure.\ref{fig:2} (e), we observe that visibility and standard deviation behave differently as a function of frame number. In fact, as the number of frames increases, average values for visibilities increase and standard deviations decrease. The data was recorded for 5000, 10 000, 20 000, and 50 000 frames for moments from the $2^{nd}$ to the $20^{th}$ orders. We conducted the experiment 10 times with 5000 frames, i.e. N=5000. Then, the visibilities were calculated by retrieving the average intensity for the same two central rows. We observed that the lowest degree of practicality is generated from 5000 frames, for which its $20^{th}$ orders visibility and standard deviation read $0.69\pm 0.17$.  Indeed, this low frame number yields the lowest visibility values and, as compared to the other frame numbers, much larger error bars. Thus, using 5000 frames fails to qualify as a practical imaging technique. In contrast, using 50 000 frames yields highly improved characteristics both for visibility and standard deviation. As compared to the values obtained with 5000 frames at the $20^{th}$ order, 50 000 frames produce a visibility improvement of $22\%$ with a 4-fold decrease in the standard deviation, which reads ($0.84 \pm 0.04$).


Moreover, 10 000 and 20 000 frames have acceptable values that certainly converge to the 50 000 frame values, thus we anticipate that increasing the frame number will probably not improve the visibility and only slightly improve the standard deviation.

\begin{figure}[htbp]
\centering
\fbox{\includegraphics[width=\linewidth]{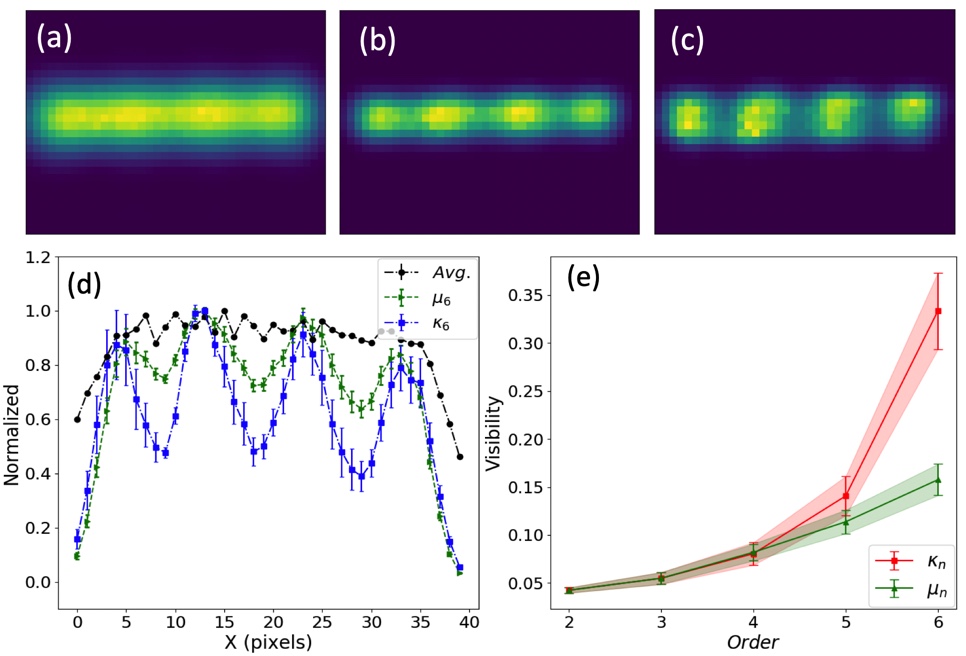}}
\caption{Color online. Comparison of cumulant and moment imaging. (a) The $2^{nd}$ order central moment $\mu_2$ image. Since the ratio of the Raleigh limit to dots separation is $50/25 = 2$, which is larger than the resolution improvement $\sqrt{2}$ that $\mu_2$ yields, the image is blurred. (b, c) for $6^{th}$ order moment and cumulant image, respectively. (d) The contrast comparison of different order images by taking two rows of pixels that are centered with the holes, and (e) the visibility and standard deviation of moment versus cumulant with 50 000 frames.}
\label{fig:3}
\end{figure}

As previously mentioned, central moments beyond the $3^{rd}$ order suffer cross-terms that worsen the resolution. Those cross-terms can effectively be eliminated by using nonlinear combinations of lower order moments. In this manner, cumulants display a significant improvement in surpassing the Rayleigh limit as compared to central moments. This can be observed in the image in figure.\ref{fig:3} (b), (c), whereby the moment and cumulant post-processing of the $6^{th}$ order is compared. The difference in contrast is highlighted in figure.\ref{fig:3} (d), as can be observed by taking two rows of pixels that are centered with the holes. Indeed though, the  $6^{th}$ order central moment $\mu_6$ shows much improvements but not as compared to the $6^{th}$ cumulant $\kappa_6$, which reads more than a two-fold improvement with respect to the moment of same order. The cumulant is computed based the nonlinear combination of moments, so the error bars increase minimally.

In addition, central moments and cumulants are compared by plotting their respective visibilities for 50,000 frames as a function of orders. $\mu_n$ and $\kappa_n$ are plotted from the $2^{nd}$  to $6^{th}$  orders with their error bars. As illustrated in figure.\ref{fig:3} (e), the average of the second and third orders are the same for central moments and cumulants, which can be explained mathematically since $\mu_n=\kappa_n$ for $n = 2,3$. The $4^{th}$ order central moment and cumulant result in very similar visibilities, since the cross-terms only contribute minimally to noise. In fact, a large difference in visibility is only observable at the $6^{th}$ order cumulant, where $\kappa_6$ yields approximately the same visibility as the $9^{th}$ order central moment with an average visibility of $\sim 0.3$, as can be seen in figure.\ref{fig:2} (d).

\section{Conclusion}

By using speckle illumination with high order correlations, we demonstrate an imaging scheme that goes beyond the sub-Rayleigh limit. We show that the object's true features can be recovered where a traditional diffraction-limited imaging method yields a completely blurred image. This is done by correlating photon counts at each pixel with two post-processing functions: moments and cumulants. The highest order moment ($n = 20$) gives the highest contrast/visibility of $0.84\pm 0.04$. In addition, and more importantly, we explore using cumulants, which, as demonstrated, show much more improvement as compared to moments starting at the $5^{th}$ order. An interesting extension of this method would be in the imaging of gray objects. Our results clearly show the capability of higher order intensity cumulants in super-resolution applications where speckles are used.  This method widens the possibilities for high order correlation imaging specifically for uses in bio-imaging and astronomy. In biomedical optics, one of standard imaging methods is laser speckle contrast imaging (LSCI), which is based on the $2^{nd}$ order correlation \cite{boas2010laser,davis2016sensitivity,briers1996laser,AMINFAR201952,ZHANG201938}. With the support of the results presented in our work, this high order correlation method could provide a competitive edge with LSCI. Moreover, speckle imaging has achieved   high resolution to identify twin stars with far less cost in time and means, which motivates speckle imaging applications in astronomy \cite{Howell2018high}.

\onecolumngrid
\appendix
\section{Correlation calculation}
\label{Correlation calculation}

Based on the experiment setup figure1(a), after the scattering, the random electric field at the $\overrightarrow{\rho}^{''}$, ground glass plane, is described by: $\left\langle E\left(\overrightarrow{\rho}_{1}^{''}\right)E\left(\overrightarrow{\rho}_{2}^{''}\right)\right\rangle =w\left(\overrightarrow{\rho}_{1}^{''}\right)\delta\left(\overrightarrow{\rho}_{1}^{''}-\overrightarrow{\rho}_{2}^{''}\right)$
, where $\left\langle ...\right\rangle $ denotes time averaging. The electric field generated from ground glass propagates to the mask, the transmission part of electric field at the $\overrightarrow{\rho}^{'}$ plane is given by $E\left(\overrightarrow{\rho}^{'}\right)=t\left(\overrightarrow{\rho}^{'}\right)\int G\left(\overrightarrow{\rho}^{'}-\overrightarrow{\rho}^{''},d^{''}\right)E\left(\overrightarrow{\rho}^{''}\right)d\overrightarrow{\rho}^{''}$
, where $G\left(\overrightarrow{\rho}^{'}-\overrightarrow{\rho}^{''},d^{''}\right)\backsimeq\frac{e^{ikd^{''}}}{i\lambda d^{''}}e^{\frac{i\pi}{\lambda d^{''}}\left(\overrightarrow{\rho}^{'}-\overrightarrow{\rho}^{''}\right)^{2}}$
 [18, 21] is the Fresnel propagator, and $t(\overrightarrow{\rho}^{'})$ is the transmission coefficient of the mask object. 

So, the  first order correlation function of transmitted field immediately after mask is given by


\begin{equation}
\begin{aligned}
 & G^{(1)}\left(\overrightarrow{\rho}_{1}^{'},\overrightarrow{\rho}_{2}^{'}\right)\\
= & \left\langle E^{*}\left(\overrightarrow{\rho}_{1}^{'}\right)E\left(\overrightarrow{\rho}_{2}^{'}\right)\right\rangle \\
= & \left\langle \int\int t^{*}\left(\overrightarrow{\rho}_{1}^{'}\right)G^{*}\left(\overrightarrow{\rho}_{1}^{'}-\overrightarrow{\rho}_{1}^{''},d^{''}\right)t\left(\overrightarrow{\rho}_{2}^{'}\right)G\left(\overrightarrow{\rho}_{2}^{'}-\overrightarrow{\rho}_{2}^{''},d^{''}\right)\right.\\
&\qquad \left.E^{*}\left(\overrightarrow{\rho}_{1}^{''}\right)E\left(\overrightarrow{\rho}_{2}^{''}\right)d\overrightarrow{\rho}_{1}^{''}d\overrightarrow{\rho}_{2}^{''}\right\rangle \\
= & t^{*}\left(\overrightarrow{\rho}_{1}^{'}\right)t\left(\overrightarrow{\rho}_{2}^{'}\right)\int G^{*}\left(\overrightarrow{\rho}_{1}^{'}-\overrightarrow{\rho}_{1}^{''},d^{''}\right)G\left(\overrightarrow{\rho}_{2}^{'}-\overrightarrow{\rho}_{1}^{''},d^{''}\right)w\left(\overrightarrow{\rho}_{1}^{''}\right)d\overrightarrow{\rho}_{1}^{''}\\
= & \frac{t^{*}\left(\overrightarrow{\rho}_{1}^{'}\right)t\left(\overrightarrow{\rho}_{2}^{'}\right)}{\left(\lambda d^{''}\right)^{2}}\int e^{\frac{-i\pi}{\lambda d^{''}}\left(\overrightarrow{\rho}_{1}^{'}-\overrightarrow{\rho}_{1}^{''}\right)^{2}}e^{\frac{i\pi}{\lambda d^{''}}\left(\overrightarrow{\rho}_{2}^{'}-\overrightarrow{\rho}_{1}^{''}\right)^{2}}w\left(\overrightarrow{\rho}_{1}^{''}\right)d\overrightarrow{\rho}_{1}^{''}\\
= & \frac{t^{*}\left(\overrightarrow{\rho}_{1}^{'}\right)t\left(\overrightarrow{\rho}_{2}^{'}\right)}{\left(\lambda d^{''}\right)^{2}}e^{\frac{-i\pi}{\lambda d^{''}}\left(\overrightarrow{\rho}_{1}^{'2}-\overrightarrow{\rho}_{2}^{'2}\right)}\int e^{\frac{2i\pi}{\lambda d^{''}}\left(\overrightarrow{\rho}_{1}^{'}-\overrightarrow{\rho}_{2}^{'}\right)\overrightarrow{\rho}_{1}^{''}}w\left(\overrightarrow{\rho}_{1}^{''}\right)d\overrightarrow{\rho}_{1}^{''}\\
= & \frac{t^{*}\left(\overrightarrow{\rho}_{1}^{'}\right)t\left(\overrightarrow{\rho}_{2}^{'}\right)}{\left(\lambda d^{''}\right)^{2}}e^{\frac{-i\pi}{\lambda d^{''}}\left(\overrightarrow{\rho}_{1}^{'2}-\overrightarrow{\rho}_{2}^{'2}\right)}w\int_{0}^{D_{S}/2}e^{\frac{2i\pi}{\lambda d^{''}}\left(\overrightarrow{\rho}_{1}^{'}-\overrightarrow{\rho}_{2}^{'}\right)\overrightarrow{\rho}_{1}^{''}}d\overrightarrow{\rho}_{1}^{''}\\
= & \frac{t^{*}\left(\overrightarrow{\rho}_{1}^{'}\right)t\left(\overrightarrow{\rho}_{2}^{'}\right)2\pi D_{s}w}{\left(\lambda d^{''}\right)^{2}}e^{\frac{-i\pi}{\lambda d^{''}}\left(\overrightarrow{\rho}_{1}^{'2}-\overrightarrow{\rho}_{2}^{'2}\right)}somb\left(\frac{2\pi D_s}{\lambda d^{''}}\left|\overrightarrow{\rho}_{1}^{'}-\overrightarrow{\rho}_{2}^{'}\right|\right)\\
\propto & t^{*}\left(\overrightarrow{\rho}_{1}^{'}\right)t\left(\overrightarrow{\rho}_{2}^{'}\right)e^{-(\frac{\lambda d^{''}}{\pi D_{S}})^2\left(\overrightarrow{\rho}_{1}^{'}-\overrightarrow{\rho}_{2}^{'}\right)^{2}}
\end{aligned}
\end{equation}

where  $somb(x)=2J_1 (x)/x$, $D_s$ is the diameter of laser beam. Here $e^{\frac{-i\pi}{\lambda d^{''}}\left(\overrightarrow{\rho}_{1}^{'2}-\overrightarrow{\rho}_{2}^{'2}\right)}$ is only a phase factor and can be neglected.
In respect, the intensity fluctuation correlation can be simplified as

\begin{equation}
\begin{aligned}
&\left\langle \delta I\left(\overrightarrow{\rho}_{1}^{'}\right)\delta I\left(\overrightarrow{\rho}_{2}^{'}\right)\right\rangle \\
& =\left\langle \left(I\left(\overrightarrow{\rho}_{1}^{'}\right)-\left\langle I\left(\overrightarrow{\rho}_{1}^{'}\right)\right\rangle \right)\left(I\left(\overrightarrow{\rho}_{2}^{'}\right)-\left\langle I\left(\overrightarrow{\rho}_{2}^{'}\right)\right\rangle \right)\right\rangle \\
& =\left\langle E^{*}\left(\overrightarrow{\rho}_{1}^{'}\right)E\left(\overrightarrow{\rho}_{1}^{'}\right)E^{*}\left(\overrightarrow{\rho}_{2}^{'}\right)E\left(\overrightarrow{\rho}_{2}^{'}\right)\right\rangle -\left\langle I\left(\overrightarrow{\rho}_{1}^{'}\right)I\left(\overrightarrow{\rho}_{2}^{'}\right)\right\rangle \\
 & =\left\langle E^{*}\left(\overrightarrow{\rho}_{1}^{'}\right)E\left(\overrightarrow{\rho}_{2}^{'}\right)\right\rangle ^{2}\\
 & \propto\left|t\left(\overrightarrow{\rho}_{1}^{'}\right)\right|^{2}\left|t\left(\overrightarrow{\rho}_{2}^{'}\right)\right|^{2}e^{-\frac{\left(\overrightarrow{\rho}_{1}^{'}-\overrightarrow{\rho}_{2}^{'}\right)^{2}}{l_{c}^{2}}}.
\end{aligned}
\end{equation}

with the coherent length  $l_{c} = \frac{\lambda d^{''}}{\sqrt{2}\pi D_{S}}$.

\section{The cumulant formulas}
\label{The cumulant formulas}

To simplify the calculation, the cumulant formulas are expressed by the moment form. The $n_{th}$ order moment image is defined by

\begin{eqnarray}
\mu_{n}\equiv M^{(n)}\left(0\right)=\left\langle \left(\sum_{i}h_{i}\delta I_{i}\right)^{n}\right\rangle,
\end{eqnarray}
where $M^{(n)}\left(\beta\right)=\left\langle e^{\beta\delta I\left(\overrightarrow{\rho}\right)}\right\rangle =\left\langle e^{\beta\sum_{i}h_{i}\delta I_{i}}\right\rangle $ is the moment-generating function. The definition of cumulant is given by 

\begin{eqnarray}
\kappa_{n}\equiv K^{(n)}\left(0\right),
\end{eqnarray}
where $K^{(n)}\left(\beta\right)=ln\left\langle e^{\beta\delta I\left(\overrightarrow{\rho}\right)}\right\rangle =ln\left\langle e^{\beta\sum_{i}h_{i}\delta I_{i}}\right\rangle $ is the cumulant-generating function. 

The moment-generating function can be written as $M(\beta)=exp(K(\beta))$. Taking $n^{th}$ order derivative in respect to $\beta$,  it reads

\begin{equation}
\begin{aligned}
M^{(n)}\left(\beta\right)=\sum_{i=0}^{n-1}\left(\begin{array}{c}
n-1\\
i
\end{array}\right)K^{(n-i)}\left(\beta\right)M^{(i)}\left(\beta\right).
\end{aligned}
\end{equation}

Let $\beta = 0$, Eq.(B3) gives

\begin{equation}
\begin{aligned}
\mu_{n} & =\sum_{i=0}^{n-1}\left(\begin{array}{c}
n-1\\
i
\end{array}\right)\kappa_{n-i}\mu_{i}\\
 & =\kappa_{n}+\sum_{i=1}^{n-1}\left(\begin{array}{c}
n-1\\
i
\end{array}\right)\kappa_{n-i}\mu_{i}, 
\end{aligned}
\end{equation}
rewriting gives the recursively relation that

\begin{equation}
\begin{aligned}
\kappa_{n}=\mu_{n}-\sum_{i=1}^{n-1}\left(\begin{array}{c}
n-1\\
i
\end{array}\right)\kappa_{n-i}\mu_{i}.
\end{aligned}
\end{equation}

The fluctuation intensity sample gives the central moments with $\mu_1=0$. Drop all terms in which $\mu_1$ appears, the first six order cumulants in form of moment are listed below:

\begin{equation}
\begin{aligned}
\kappa_{1} & =\mu_{1}\\
\kappa_{2} & =\mu_{2}\\
\kappa_{3} & =\mu_{3}\\
\kappa_{4} & =\mu_{4}-3\mu_{2}^{2}\\
\kappa_{5} & =\mu_{5}-10\mu_{3}\mu_{2}\\
\kappa_{6} & =\mu_{6}-15\mu_{4}\mu_{2}-10\mu_{3}^{2}+30\mu_{2}^{3}.
\end{aligned}
\end{equation}

\twocolumngrid

\bibliographystyle{unsrt}
\bibliography{main}

\begin{thebibliography}{10}

\bibitem{goodman2007speckle}
Joseph~W Goodman.
\newblock {\em Speckle phenomena in optics: theory and applications}.
\newblock Roberts and Company Publishers, 2007.

\bibitem{goodman1976some}
Joseph~W Goodman.
\newblock Some fundamental properties of speckle.
\newblock {\em JOSA}, 66(11):1145--1150, 1976.

\bibitem{dunn2001dynamic}
Andrew~K Dunn, Hayrunnisa Bolay, Michael~A Moskowitz, and David~A Boas.
\newblock Dynamic imaging of cerebral blood flow using laser speckle.
\newblock {\em Journal of Cerebral Blood Flow \& Metabolism}, 21(3):195--201,
  2001.

\bibitem{Dainty1975laser}
J.C. Dainty.
\newblock {\em Laser Speckle and Related Phenomena}.
\newblock Springer-Verlag Berlin Heidelberg, 1975.

\bibitem{gatti2008progress}
A~Gatti, E~Brambilla, and L~Lugiato.
\newblock Progress in optics vol 51 ed e wolf, 2008.

\bibitem{Devaux_2016}
Fabrice Devaux, Kien~Phan Huy, S{\'{e}}verine Denis, Eric Lantz, and
  Paul-Antoine Moreau.
\newblock Temporal ghost imaging with pseudo-thermal speckle light.
\newblock {\em Journal of Optics}, 19(2):024001, dec 2016.

\bibitem{dogariu2003propagation}
Aristide Dogariu and Stefan Amarande.
\newblock Propagation of partially coherent beams: turbulence-induced
  degradation.
\newblock {\em Optics letters}, 28(1):10--12, 2003.

\bibitem{shirai2003mode}
Tomohiro Shirai, Aristide Dogariu, and Emil Wolf.
\newblock Mode analysis of spreading of partially coherent beams propagating
  through atmospheric turbulence.
\newblock {\em JOSA A}, 20(6):1094--1102, 2003.

\bibitem{siviloglou2007observation}
GA~Siviloglou, J~Broky, Aristide Dogariu, and DN~Christodoulides.
\newblock Observation of accelerating airy beams.
\newblock {\em Physical Review Letters}, 99(21):213901, 2007.

\bibitem{baleine2006correlated}
Erwan Baleine, Aristide Dogariu, and Girish~S Agarwal.
\newblock Correlated imaging with shaped spatially partially coherent light.
\newblock {\em Optics letters}, 31(14):2124--2126, 2006.

\bibitem{gatti2004ghost}
Alessandra Gatti, Enrico Brambilla, Morten Bache, and Luigi~A Lugiato.
\newblock Ghost imaging with thermal light: comparing entanglement and
  classicalcorrelation.
\newblock {\em Physical review letters}, 93(9):093602, 2004.

\bibitem{valencia2005two}
Alejandra Valencia, Giuliano Scarcelli, Milena D’Angelo, and Yanhua Shih.
\newblock Two-photon imaging with thermal light.
\newblock {\em Physical review letters}, 94(6):063601, 2005.

\bibitem{lahiri2009determination}
Mayukh Lahiri, Emil Wolf, David~G Fischer, and Tomohiro Shirai.
\newblock Determination of correlation functions of scattering potentials of
  stochastic media from scattering experiments.
\newblock {\em Physical review letters}, 102(12):123901, 2009.

\bibitem{arecchi1965measurement}
Fortunato~T Arecchi.
\newblock Measurement of the statistical distribution of gaussian and laser
  sources.
\newblock {\em Physical Review Letters}, 15(24):912, 1965.

\bibitem{classen2017superresolution}
Anton Classen, Joachim von Zanthier, Marlan~O Scully, and Girish~S Agarwal.
\newblock Superresolution via structured illumination quantum correlation
  microscopy.
\newblock {\em Optica}, 4(6):580--587, 2017.

\bibitem{oh2013sub}
Joo-Eon Oh, Young-Wook Cho, Giuliano Scarcelli, and Yoon-Ho Kim.
\newblock Sub-rayleigh imaging via speckle illumination.
\newblock {\em Optics letters}, 38(5):682--684, 2013.

\bibitem{wang2015spatial}
Yunlong Wang, Feiran Wang, Ruifeng Liu, Dongxu Chen, Hong Gao, Pei Zhang, and
  Fuli Li.
\newblock Spatial sub-rayleigh imaging analysis via speckle laser illumination.
\newblock {\em Optics letters}, 40(22):5323--5326, 2015.

\bibitem{smith2018turbulence}
Thomas~A Smith and Yanhua Shih.
\newblock Turbulence-free double-slit interferometer.
\newblock {\em Physical review letters}, 120(6):063606, 2018.

\bibitem{sprigg2016super}
Jane Sprigg, Tao Peng, and Yanhua Shih.
\newblock Super-resolution imaging using the spatial-frequency filtered
  intensity fluctuation correlation.
\newblock {\em Scientific reports}, 6:38077, 2016.

\bibitem{dertinger2009fast}
Thomas Dertinger, Ryan Colyer, Gopal Iyer, Shimon Weiss, and J{\"o}rg
  Enderlein.
\newblock Fast, background-free, 3d super-resolution optical fluctuation
  imaging (sofi).
\newblock {\em Proceedings of the National Academy of Sciences},
  106(52):22287--22292, 2009.

\bibitem{vigoren2018optical}
Andrew Vigoren and James~M Zavislan.
\newblock Optical sectioning enhancement using higher-order moment signals in
  random speckle-structured illumination microscopy.
\newblock {\em JOSA A}, 35(3):474--479, 2018.

\bibitem{oppel2014directional}
S~Oppel, R~Wiegner, GS~Agarwal, and J~von Zanthier.
\newblock Directional superradiant emission from statistically independent
  incoherent nonclassical and classical sources.
\newblock {\em Physical review letters}, 113(26):263606, 2014.

\bibitem{chan2009high}
Kam Wai~Clifford Chan, Malcolm~N O'Sullivan, and Robert~W Boyd.
\newblock High-order thermal ghost imaging.
\newblock {\em Optics letters}, 34(21):3343--3345, 2009.

\bibitem{shih2016introduction}
Yanhua Shih.
\newblock {\em An introduction to quantum optics: photon and biphoton physics}.
\newblock CRC press, 2016.

\bibitem{boas2010laser}
David~A Boas and Andrew~K Dunn.
\newblock Laser speckle contrast imaging in biomedical optics.
\newblock {\em Journal of biomedical optics}, 15(1):011109, 2010.

\bibitem{davis2016sensitivity}
Mitchell~A Davis, Louis Gagnon, David~A Boas, and Andrew~K Dunn.
\newblock Sensitivity of laser speckle contrast imaging to flow perturbations
  in the cortex.
\newblock {\em Biomedical optics express}, 7(3):759--775, 2016.

\bibitem{briers1996laser}
J~David Briers and Sian Webster.
\newblock Laser speckle contrast analysis (lasca): a nonscanning, full-field
  technique for monitoring capillary blood flow.
\newblock {\em Journal of biomedical optics}, 1(2):174--180, 1996.

\bibitem{AMINFAR201952}
AmirHessam Aminfar, Nami Davoodzadeh, Guillermo Aguilar, and Marko Princevac.
\newblock Application of optical flow algorithms to laser speckle imaging.
\newblock {\em Microvascular Research}, 122:52 -- 59, 2019.

\bibitem{ZHANG201938}
Ru~Zhang, Lipei Song, Jiachun Xu, Xiaoying An, Weiming Sun, Xing Zhao, Zhen
  Zhou, and Lingling Chen.
\newblock Laser speckle imaging for blood flow based on pixel resolved
  zero-padding auto-correlation coefficient distribution.
\newblock {\em Optics Communications}, 439:38 -- 46, 2019.

\bibitem{Howell2018high}
Elliott~Horch Steve~Howell.
\newblock High-resolution speckle imaging.
\newblock {\em Physics Today}, 71(11):78--79, 2018.

\end{thebibliography}


\end{document}